\documentclass[preprintnumbers,amsmath,amssymbm,prd]{revtex4}
\usepackage{epsfig}
\usepackage{graphicx}
\begin{document}
\title{On the stability of highly-charged Reissner-Nordstr\"om black holes to charged scalar perturbations}
\author{Shahar Hod}
\affiliation{The Ruppin Academic Center, Emeq Hefer 40250, Israel}
\affiliation{ } \affiliation{The Hadassah Institute, Jerusalem
91010, Israel}
\date{\today}

\begin{abstract}
\ \ \ The stability of Reissner-Nordstr\"om black holes under the
influence of {\it neutral} perturbation fields was proved by
Moncrief four decades ago. However, the superradiant scattering
phenomenon, which characterizes the dynamics of {\it charged}
bosonic fields in these charged black-hole spacetimes, imposes a
greater and non-trivial threat on their stability. According to this
well-known phenomenon, integer-spin charged fields interacting with
a Reissner-Nordstr\"om black hole can be amplified (gain energy) by
extracting some of the black-hole Coulomb energy. If, in addition to
being electrically charged, the incident bosonic fields also possess
non-zero rest masses, then the mutual gravitational attraction
between the central black hole and the fields may prevent the
extracted energy and electric charge from escaping to infinity. One
may suspect that the physical mechanism of superradiant
amplification of charged bosonic fields in the charged
Reissner-Nordstr\"om black-hole spacetime, when combined with the
confinement mechanism provided by the mutual gravitational
attraction between the black hole and the massive fields, may lead
to a superradiant instability of the Reissner-Nordstr\"om black-hole
spacetime. (This suspicion is mainly based on our experience with
rotating Kerr black holes, which are known to be characterized by an
analogous superradiant instability when coupled to massive bosonic
fields). However, in this paper we show that, for highly-charged
Reissner-Nordstr\"om black holes in the charge interval
$8/9<{(Q/M)}^2<1$, the two physical mechanisms which are required in
order to trigger the superradiant instability phenomenon in the
black-hole spacetime [namely: (1) the superradiant amplification of
incident charged scalar fields by the charged black hole, and (2)
the existence of a binding potential well in the black-hole exterior
region which prevents the extracted energy and electric charge from
escaping to infinity] {\it cannot} operate simultaneously. In
particular, we shall prove that, in the superradiant regime, there
is no trapping potential well in the black-hole exterior region. Our
stability results for the highly-charged Reissner-Nordstr\"om black
holes in the regime $8/9<(Q/M)^2<1$, combined with former analytical
studies which explored the stability of these charged black holes in
the complementary regime ${(Q/M)}^2\leq 8/9$, establish the fact
that charged Reissner-Nordstr\"om black holes are stable to charged
massive scalar perturbations in the entire parameter space.
\end{abstract}
\bigskip
\maketitle


\section{Introduction}

Rotating Kerr black holes are known to be super-radiantly unstable
under the influence of co-rotating massive bosonic fields
\cite{Zel,PressTeu1,Ins2,Hodstat}. It has recently been shown that,
for a scalar field of mass $\mu$ and azimuthal harmonic index $m$,
the instability regime is bounded by the relation
\cite{Hodbou,Hodcq,Noteunt}
\begin{equation}\label{Eq1}
0<\mu<m\Omega_{\text{H}}\  ,
\end{equation}
where $\Omega_{\text{H}}$ is the angular velocity of the black-hole
horizon \cite{Notesat}.

The instability of the composed Kerr-massive-bosonic-field system is
a direct consequence of the superradiance phenomenon which
characterizes the dynamics of integer-spin fields in spinning
black-hole spacetimes \cite{Zel,PressTeu1,Ins2,Hodstat,Ins1}. In
particular, it is well established that a bosonic field mode of
frequency $\omega$ impinging upon a spinning Kerr black hole is
amplified (extracts rotational energy and angular momentum from the
black hole) if the superradiance condition \cite{Zel,PressTeu1}
\begin{equation}\label{Eq2}
\omega<m\Omega_{\text{H}}\
\end{equation}
is satisfied.

While the physical mechanism of superradiant amplification
\cite{Zel,PressTeu1,Ins2,Hodstat,Ins1} is required in order to
extracts rotational energy from the spinning Kerr black hole, an
additional physical mechanism is required in order to trigger a
genuine exponentially growing instability in the black-hole
spacetime. In particular, for the instability to exist, one should
prevent the bosonic fields from radiating their energy to infinity.
For the case of {\it massive} bosonic fields propagating in the
black-hole spacetime, it is the attractive gravitational interaction
between the central black hole and the massive fields which provides
a natural confinement mechanism, thus preventing the extracted
rotational energy from escaping to infinity. In particular, it is
well established \cite{Zel,PressTeu1,Ins2,Hodstat} that field modes
whose conserved frequencies are bounded from above by
\begin{equation}\label{Eq3}
\omega^2<\mu^2
\end{equation}
are prevented from escaping to infinity [see Eq. (\ref{Eq15})
below].

A similar physical mechanism, which involves the superradiant
amplification of {\it charged} bosonic fields, threatens the
stability of {\it charged} Reissner-Nordstr\"om black holes. In this
case, the superradiant scattering phenomenon (that is, the
amplification of incident charged bosonic fields) occurs for
frequencies in the regime \cite{Bekch}
\begin{equation}\label{Eq4}
\omega<q\Phi_{\text{H}}\  ,
\end{equation}
where $q$ and $\Phi_{\text{H}}=Q/r_+$ are respectively the charge
coupling constant of the incident scalar field and the electric
potential of the central Reissner-Nordstr\"om black hole. (Here $Q$
and $r_+$ are respectively the electric charge and horizon radius of
the black hole \cite{NoteQ}).

The scattering of charged bosonic fields in the superradiant regime
(\ref{Eq4}) is characterized by the extraction of Coulomb energy and
electric charge from the charged Reissner-Nordstr\"om (RN) black
hole \cite{Bekch}. It is thus natural to suspect that the
superradiant phenomenon, which characterizes the dynamics of charged
massive bosonic fields in the charged RN black-hole spacetime,
together with the natural confinement mechanism (\ref{Eq3}) provided
by the attractive gravitational interaction between the central
black hole and the massive fields, may lead to an instability of the
charged RN spacetime \cite{Noteme}.

The suspicion that the charged RN black-hole spacetime may be
unstable to charged massive scalar perturbations is based on the
similarity between the physical phenomenon of superradiant
amplification of charged bosonic fields in the charged RN black-hole
spacetime and the physical phenomenon of superradiant amplification
of rotating massive bosonic fields in the rotating Kerr black-hole
spacetime. As discussed above, in the later case the superradiant
amplification of massive scalar perturbations is known to produce
exponentially growing instabilities in the Kerr black-hole
spacetime.

However, in \cite{Hodnbt} we have proved that charged RN black holes
in the regime
\begin{equation}\label{Eq5}
(Q/M)^2\leq {8\over 9}
\end{equation}
are actually {\it stable} under the influence of charged scalar
perturbations \cite{Notenbt2}. The main goal of the present paper is
to analyze the stability under charged scalar perturbations of {\it
highly-charged} RN black holes in the complementary regime
\begin{equation}\label{Eq6}
{8\over 9}<(Q/M)^2<1\  .
\end{equation}
Below we shall provide evidence for the stability of the composed
RN-charged-scalar-field system in the regime (\ref{Eq6}). In
particular, we shall show that, for highly-charged RN black holes in
the regime (\ref{Eq6}), the two physical mechanisms which are
required in order to trigger the superradiant instability phenomenon
[namely: (1) the superradiant amplification of the charged bosonic
fields, and (2) the existence of a binding potential well in the
black-hole exterior region which prevents the extracted energy and
electric charge from escaping to infinity] {\it cannot} operate
simultaneously. Our results thus support the stability of these
highly-charged RN black holes.

\section{Description of the system}

The physical system we shall analyze is composed of a massive
charged scalar field which is linearly coupled to a charged
Reissner-Nordstr\"om black hole. The RN black-hole spacetime is
described by the line element \cite{Chan}
\begin{equation}\label{Eq7}
ds^2=-{{\Delta(r)}\over{r^2}}dt^2+{{r^2}\over{\Delta(r)}}dr^2+r^2(d\theta^2+\sin^2\theta
d\phi^2)\ ,
\end{equation}
where \cite{Noteunt}
\begin{equation}\label{Eq8}
\Delta\equiv r^2-2Mr+Q^2\  .
\end{equation}
Here $M$ and $Q$ are respectively the mass and electric charge of
the black hole, and $r$ is the radial areal coordinate. The
black-hole (event and inner) horizons are determined by the zeroes
of $\Delta(r)$:
\begin{equation}\label{Eq9}
r_{\pm}=M\pm \sqrt{M^2-Q^2}\  .
\end{equation}

The dynamics of the charged massive scalar field $\Psi$ in the
charged RN black-hole spacetime is described by the Klein-Gordon
wave equation \cite{Stro,HodPirpam,HodCQG2,Hodpla}
\begin{equation}\label{Eq10}
[(\nabla^\nu-iqA^\nu)(\nabla_{\nu}-iqA_{\nu}) -\mu^2]\Psi=0\  .
\end{equation}
Here $\mu$ and $q$ are respectively the mass and charge coupling
constant of the field \cite{Notemq}, and
$A_{\nu}=-\delta_{\nu}^{0}{Q/r}$ is the electromagnetic potential of
the black hole. It proves useful to decompose the scalar field in
the form
\begin{equation}\label{Eq11}
\Psi(t,r,\theta,\phi)=\sum_{lm}e^{im\phi}S_{lm}(\theta)\Delta^{-1/2}\psi_{lm}(r)e^{-i\omega
t}\ ,
\end{equation}
where $\omega$ is the conserved frequency of the field mode, and
$\{l,m\}$ are respectively the spherical harmonic index and the
azimuthal harmonic index of the mode \cite{Noteom}. Remembering that
any instability must set in via a real-frequency mode
\cite{PressTeu1,Ins2,Hodstat,Hartle}, we shall consider here
marginally-stable field modes with $\omega_I=0$.

Substituting the field decomposition (\ref{Eq11}) into the
Klein-Gordon wave equation (\ref{Eq10}), one finds that the radial
function $\psi$ and the angular function $S$ are determined by two
ordinary differential equations which are coupled by a separation
constant $K_l=l(l+1)$ \cite{Heun,Flam,Abram}. The radial equation
for $\psi$ takes the form of a Schr\"odinger-like wave equation
\cite{Stro,HodPirpam,HodCQG2,Hodnbt}
\begin{equation}\label{Eq12}
{{d^2\psi}\over{dr^2}}+(\omega^2-V)\psi=0\  ,
\end{equation}
where
\begin{equation}\label{Eq13}
\omega^2-V\equiv{{U+M^2-Q^2}\over{\Delta^2}}\
\end{equation}
with
\begin{equation}\label{Eq14}
U\equiv(\omega r^2-qQr)^2-\Delta[\mu^2r^2+l(l+1)]\  .
\end{equation}
Note that the bound-state resonances of the scalar field in the
black-hole spacetime are characterized by exponentially decaying
eigenfunctions at large distances from the black hole
\cite{Ins2,Hodstat}:
\begin{equation}\label{Eq15}
\psi(r\to\infty) \sim re^{-\sqrt{\mu^2-\omega^2}r}\ \ \ \text{with}\
\ \ \omega^2<\mu^2\  .
\end{equation}

\section{No bound-state resonances in the superradiant regime}

As discussed in the Introduction, two distinct physical mechanisms
are required to operate simultaneously in order to trigger a
superradiant instability in the black-hole spacetime:
\begin{itemize}
\item{The extraction of Coulomb energy from the charged black hole due to
the physical mechanism of superradiant amplification of charged
bosonic fields in the superradiant regime $\omega<qQ/r_+$ [see Eq.
(\ref{Eq4})].}
\item{The existence of a binding potential well in the
black-hole exterior region which prevents the extracted energy from
escaping to infinity. This trapping potential is required in order
to support the bound-state resonances of the charged scalar fields
in the black-hole exterior region. In particular, in order to
support these bound-state scalar configurations, the potential well
must be separated from the black-hole horizon by a potential
barrier.}
\end{itemize}
We shall prove below that these two physical mechanisms (that is,
superradiant amplification and trapping of the fields) cannot
operate simultaneously in the regime (\ref{Eq6}) of highly-charged
RN black holes.

Specifically, we shall prove below that, in the superradiant regime
(\ref{Eq4}), there are no bound-state resonances of the charged
scalar fields in the black-hole exterior region [that is, we shall
prove that, for highly-charged RN black holes in the regime
(\ref{Eq6}), there are no trapping potential wells in the black-hole
exterior region which are separated from the black-hole horizon by a
potential barrier]. To that end, we shall now study the properties
of the effective potential $V(r;M,Q,\mu,q,\omega,l)$ that appears in
the Schr\"odinger-like radial wave equation (\ref{Eq12}).

\subsection{The gradient of the effective radial potential}

The gradient of the effective radial potential $V(r)$ can be
expressed as a $4$th order polynomial function \cite{Hodnbt}
\begin{equation}\label{Eq16}
-{{\Delta^3}\over{2}}V'(z;M,Q,\mu,q,\omega,l)=az^4+bz^3+cz^2+dz+e\ ,
\end{equation}
where
\begin{equation}\label{Eq17}
z\equiv r-r_-\  .
\end{equation}
The expansion coefficients in (\ref{Eq16}) are given by
\cite{Notehnb}
\begin{equation}\label{Eq18}
a=M\mu^2+Qq\omega-2M\omega^2\  ,
\end{equation}
\begin{equation}\label{Eq19}
b=(4Mr_--2M^2-Q^2)\mu^2+(-8Mr_-+2Q^2)\omega^2+2Q(M+2r_-)q\omega-Q^2q^2+l(l+1)\
,
\end{equation}
\begin{equation}\label{Eq20}
c=-3r^2_-(r_+-M)\mu^2+3r^3_-({{qQ}\over{r_-}}-\omega)(2\omega-{{qQ}\over{r_-}})-3(r_+-M)l(l+1)\
,
\end{equation}
and
\begin{equation}\label{Eq21}
e=2r^4_-(r_+-M)(\omega-{{qQ}\over{r_-}})^2+2(r_+-M)^3\ .
\end{equation}

Note that the condition (\ref{Eq3}) for the existence of bound-state
scalar resonances in the black-hole exterior region, together with
the condition (\ref{Eq4}) for the existence of the superradiant
amplification phenomenon of the trapped bosonic modes, imply that
the mode frequency $\omega$ is restricted to the interval
\begin{equation}\label{Eq22}
0\leq\omega<\text{min}\big\{{{qQ}\over{r_+}},\mu\big\}\  .
\end{equation}

\subsection{The signs of the coefficients $\{a,b,c,e\}$}

In order to analyze the spatial behavior of the effective radial
potential $V(r)$, we shall first analyze the signs of the
coefficients $\{a,b,c,e\}$ that appear in the gradient equation
(\ref{Eq16}).

It was proved in \cite{Hodnbt} that, in the frequency interval
(\ref{Eq22}), the coefficient $a$ is positive definite in the entire
range $0\leq (Q/M)^2\leq1$ \cite{Noteaa}:
\begin{equation}\label{Eq23}
a>0\  .
\end{equation}
In addition, inspection of Eq. (\ref{Eq21}) reveals that the
coefficient $e$ is positive definite in the entire range $0\leq
(Q/M)^2\leq1$ \cite{Hodnbt}:
\begin{equation}\label{Eq24}
e>0\  .
\end{equation}

We shall now prove that, for highly-charged Reissner-Nordstr\"om
black holes in the charge interval [see Eq. (\ref{Eq6})]
\begin{equation}\label{Eq25}
{8\over 9}<{x}^2<1\ \ \ \ ; \ \ \ \ x\equiv Q/M\  ,
\end{equation}
(at least) one of the inequalities
\begin{equation}\label{Eq26}
b>0\ \ \ \ ; \ \ \ \ c<0\  ,
\end{equation}
holds true. In particular, we shall henceforth assume that
\cite{Noteqa}
\begin{equation}\label{Eq27}
c>0\  ,
\end{equation}
and prove that this inequality necessarily implies the inequality
\begin{equation}\label{Eq28}
b>0
\end{equation}
in the frequency interval (\ref{Eq22}).

Note that a necessary condition for the validity of the inequality
$c(\omega)>0$ is given by
$({{qQ}/{r_-}}-\omega)(2\omega-{{qQ}/{r_-}})>0$ [see Eq.
(\ref{Eq20})], or equivalently by
${{qQ}/{2r_-}}<\omega<{{qQ}/{r_-}}$. Taking cognizance of
(\ref{Eq22}) and using the fact that $r_-<r_+$, one realizes that in
the regime (\ref{Eq27}) the conserved mode frequency is bounded by
\cite{Notemuqq}
\begin{equation}\label{Eq29}
{{qQ}\over{2r_-}}<\omega<\text{min}\big\{{{qQ}\over{r_+}},\mu\big\}\
.
\end{equation}

There are two distinct cases which should be analyzed separately:

Case (A): Charged Reissner-Nordstr\"om black holes in the interval
\begin{equation}\label{Eq30}
8/9<x^2\leq 4\sqrt{3}-6\  .
\end{equation}
This charge interval corresponds to the inequality
$4Mr_--2M^2-Q^2\leq 0$ for the coefficient of $\mu^2$ in
(\ref{Eq19}), in which case one can obtain a lower bound on the
value of the coefficient $b(\omega)$ by substituting into
(\ref{Eq19}) the maximally allowed value of the mass parameter
$\mu^2$. This value is given by the inequality [see Eq. (\ref{Eq20})
with $c>0$]
\begin{equation}\label{Eq31}
0<
3r^2_-(r_+-M)\mu^2<3r^3_-({{qQ}\over{r_-}}-\omega)(2\omega-{{qQ}\over{r_-}})-3(r_+-M)l(l+1)\
.
\end{equation}

Substituting (\ref{Eq31}) into (\ref{Eq19}), one finds
\begin{equation}\label{Eq32}
b>I(\omega;x)+{{2\sqrt{1-x^2}}\over{(1-\sqrt{1-x^2})^2}}\cdot
l(l+1)\ ,
\end{equation}
where
\begin{eqnarray}\label{Eq33}
I(\omega;x)\equiv
{{2x^2-4+4\sqrt{1-x^2}}\over{\sqrt{1-x^2}}}\cdot(M\omega)^2+
{{x^2+2-6\sqrt{1-x^2}}\over{\sqrt{1-x^2}}}\cdot qQM\omega
+{{3\sqrt{1-x^2}-1}\over{\sqrt{1-x^2}(1-\sqrt{1-x^2})}}\cdot (qQ)^2\  .
\end{eqnarray}

Note that $2x^2-4+4\sqrt{1-x^2}<0$ in the entire range $0<x^2\leq
1$, which implies that the dependence of $I(\omega)$ on $\omega$ is
in the form of a convex parabola. Thus, $I(\omega)$ is minimized at
the boundaries of the frequency interval (\ref{Eq29}). Substituting
$\omega\to qQ/2r_-$ into (\ref{Eq33}), one finds
\begin{equation}\label{Eq34}
I(\omega\to qQ/2r_-;x)={1\over 2}(qQ)^2>0\  ,
\end{equation}
which obviously has a positive definite value. Substituting into
(\ref{Eq33}) $\omega\to qQ/r_+$ for the case $qQ/r_+\leq\mu$, one
finds
\begin{equation}\label{Eq35}
I(\omega\to
qQ/r_+;x)={{16(1-\sqrt{1-x^2})+2x^2(5\sqrt{1-x^2}-7)-x^4}\over{x^4}}\cdot
(qQ)^2>0\  ,
\end{equation}
where the inequality sign in (\ref{Eq35}) refers to RN black holes
in the entire range $0\leq x^2\leq 1$. This implies, in particular,
that the expression (\ref{Eq35}) is a positive definite function in
the charge interval (\ref{Eq30}). Substituting into (\ref{Eq33})
$\omega\to\mu$ for the case $qQ/2r_-<\mu<qQ/r_+$ \cite{Notemuqq},
one finds \cite{Notecon}
\begin{equation}\label{Eq36}
I(\omega\to\mu;x)>\text{min}\{I(\omega\to qQ/2r_-;x),I(\omega\to
qQ/r_+;x)\}>0\  .
\end{equation}
We therefore conclude that charged Reissner-Nordstr\"om black holes
in the charge interval (\ref{Eq30}) are characterized by (at least)
one of the relations [see Eqs. (\ref{Eq27}), (\ref{Eq32}),
(\ref{Eq34}), (\ref{Eq35}), and (\ref{Eq36})]
\begin{equation}\label{Eq37}
b\geq I>0\ \ \ \ ; \ \ \ \ c<0\  .
\end{equation}

Case (B): Charged Reissner-Nordstr\"om black holes in the interval
\begin{equation}\label{Eq38}
4\sqrt{3}-6<x^2<1\  .
\end{equation}
This charge interval corresponds to the inequality
$4Mr_--2M^2-Q^2>0$ for the coefficient of $\mu^2$ in (\ref{Eq19}).
Remembering that $\mu^2>\omega^2$ [see (\ref{Eq15})], one can obtain
a lower bound on the value of the coefficient $b(\omega)$ by
replacing $\mu\to\omega$ in (\ref{Eq19}). One then finds
\begin{equation}\label{Eq39}
b>J(\omega;x)+l(l+1)\ ,
\end{equation}
where
\begin{equation}\label{Eq40}
J(\omega;x)\equiv
(x^2+4\sqrt{1-x^2}-6)\cdot(M\omega)^2+2(3-2\sqrt{1-x^2})\cdot
qQM\omega-(qQ)^2\ .
\end{equation}

Note that $x^2+4\sqrt{1-x^2}-6<0$ in the entire range $0\leq x^2\leq
1$, which implies that the dependence of $J(\omega)$ on $\omega$ is
in the form of a convex parabola. Thus, $J(\omega)$ is minimized at
the boundaries of the frequency interval (\ref{Eq29}). Substituting
$\omega\to qQ/2r_-$ into (\ref{Eq40}), one finds
\begin{equation}\label{Eq41}
J(\omega\to
qQ/2r_-;x)={{6-3x^2-8\sqrt{1-x^2}}\over{4(1-\sqrt{1-x^2})^2}}\cdot(qQ)^2>0\
,
\end{equation}
where the inequality sign in (\ref{Eq41}) refers to RN black holes
in the regime $x^2>(8\sqrt{7}-14)/9$. In particular, one finds that
the expression (\ref{Eq41}) is a positive definite function in the
entire charge interval (\ref{Eq38}) [note that
$(8\sqrt{7}-14)/9<4\sqrt{3}-6$]. Substituting into (\ref{Eq40})
$\omega\to qQ/r_+$ for the case $qQ/r_+\leq\mu$, one finds
\begin{equation}\label{Eq42}
J(\omega\to
qQ/r_+;x)={{6x^2+4\sqrt{1-x^2}-6}\over{(1+\sqrt{1-x^2})^2}}\cdot(qQ)^2>0\
,
\end{equation}
where the inequality sign in (\ref{Eq42}) refers to RN black holes
in the regime $5/9<x^2<1$. In particular, one finds that the
expression (\ref{Eq42}) is a positive definite function in the
entire charge interval (\ref{Eq38}) [note that $5/9<4\sqrt{3}-6$].
Substituting into (\ref{Eq40}) $\omega\to\mu$ for the case
$qQ/2r_-<\mu<qQ/r_+$ \cite{Notemuqq}, one finds \cite{Notecon}
\begin{equation}\label{Eq43}
J(\omega\to \mu;x)>\text{min}\{J(\omega\to qQ/2r_-;x),J(\omega\to
qQ/r_+;x)\}>0\  .
\end{equation}
We therefore conclude that charged RN black holes in the charge
interval (\ref{Eq38}) are characterized by (at least) one of the
relations [see Eqs. (\ref{Eq27}), (\ref{Eq39}), (\ref{Eq41}),
(\ref{Eq42}), and (\ref{Eq43})]
\begin{equation}\label{Eq44}
b\geq J>0\ \ \ \ ; \ \ \ \ c<0\  .
\end{equation}

\subsection{The spatial behavior of the effective radial potential}

We shall now analyze the spatial behavior of the effective radial
potential $V(r)$ that appears in the Schr\"odinger-like radial
equation (\ref{Eq12}). In particular, we shall determine the signs
of the four roots $\{z_1,z_2,z_3,z_4\}$ which characterize the
gradient equation [see Eq. (\ref{Eq16})]
\begin{equation}\label{Eq45}
V'(z)=0\  .
\end{equation}

Taking cognizance of the inequality (\ref{Eq23}), one finds [see Eq.
(\ref{Eq17})]
\begin{equation}\label{Eq46}
V'(r\to\infty)\to 0^-\
\end{equation}
for the asymptotic behavior of the effective radial potential. In
addition, from Eqs. (\ref{Eq13}) and (\ref{Eq14}) one finds that the
effective radial potential $V(r)$ is characterized by the following
two properties:
\begin{equation}\label{Eq47}
V(r\to r_+)\to -\infty\  ,
\end{equation}
and
\begin{equation}\label{Eq48}
V(r\to r_-)\to -\infty\  .
\end{equation}

Taking cognizance of Eqs. (\ref{Eq46}) and (\ref{Eq47}), one
concludes that the effective potential $V(r)$ is characterized by at
least one maximum point in the exterior black-hole region $r>r_+$
\cite{Hodnbt}. We denote that maximum point by $z_4$, where
\begin{equation}\label{Eq49}
z_4>0\  .
\end{equation}
In addition, taking cognizance of Eqs. (\ref{Eq47}) and
(\ref{Eq48}), one concludes that the effective potential $V(r)$ has
at least one maximum point which is located in the interval
$r_-<r<r_+$ between the two horizons of the charged black hole
\cite{Hodnbt}. We denote that maximum point by $z_3$, where
\begin{equation}\label{Eq50}
z_4>z_3>0\  .
\end{equation}

Taking cognizance of Eqs. (\ref{Eq23}) and (\ref{Eq24}), and using
the well-known relation
\begin{equation}\label{Eq51}
z_1\cdot z_2\cdot z_3\cdot z_4={e\over a}\  ,
\end{equation}
for the four roots of a quartic equation, one finds
\begin{equation}\label{Eq52}
z_1\cdot z_2\cdot z_3\cdot z_4>0\  .
\end{equation}
Likewise, taking cognizance of Eqs. (\ref{Eq23}), (\ref{Eq37}), and
(\ref{Eq44}), and using the well-known relations
\begin{equation}\label{Eq53}
z_1+z_2+z_3+z_4=-{b\over a}\
\end{equation}
and
\begin{equation}\label{Eq54}
z_1\cdot z_2+z_1\cdot z_3+z_1\cdot z_4+z_2\cdot z_3+z_2\cdot
z_4+z_3\cdot z_4={c\over a}\
\end{equation}
for the four roots of a quartic equation, one concludes that
highly-charged RN black holes in the charge interval (\ref{Eq6}) are
characterized by (at least) one of the following relations:
\begin{equation}\label{Eq55}
z_1+z_2+z_3+z_4<0\ \ \ \ ; \ \ \ \ z_1\cdot z_2+z_1\cdot
z_3+z_1\cdot z_4+z_2\cdot z_3+z_2\cdot z_4+z_3\cdot z_4<0\ .
\end{equation}

Finally, taking cognizance of Eqs. (\ref{Eq50}), (\ref{Eq52}), and
(\ref{Eq55}), one concludes that $V'(z)$ is characterized by two
negative roots (we denote them by $z_1$ and $z_2$, where $z_1\leq
z_2<0$) and two positive roots ($0<z_3<z_4$).

Our analysis thus reveals that, for bound states of the scalar
fields [characterized by $\omega^2<\mu^2$, see (\ref{Eq15})] in the
superradiant regime [characterized by $\omega<qQ/r_+$, see
(\ref{Eq4})], the gradient $V'(z)$ of the effective potential that
appears in the Schr\"odinger-like radial equation (\ref{Eq12}) is
characterized by two positive roots and two negative roots. The
physical root $z_4>0$ (or equivalently, $r_4>r_+$) corresponds to a
{\it maximum} point of the effective radial potential $V(r)$ in the
black-hole exterior region. The negative roots $\{z_1,z_2\}$ and the
positive root $z_3$ correspond to three roots of $V'(r)$ in the
black-hole interior region $r<r_+$.

We have therefore proved that, in the superradiant regime
(\ref{Eq4}), the effective radial potential $V(r)$ is characterized
by the existence of only one {\it maximum} point (and {\it no}
minima at all) in the black-hole exterior region. This implies that,
in the superradiant regime (\ref{Eq4}), there is {\it no} binding
potential well in the physical region $r>r_+$ which is separated
from the black-hole horizon by a potential barrier. One therefore
concludes that, for highly-charged RN black holes in the charge
interval $8/9<(Q/M)^2<1$, there are no bound-state resonances (with
$\omega^2<\mu^2$) of the charged massive scalar fields in the
superradiant regime $\omega<qQ/r_+$.

\section{Summary}

Motivated by the well-known superradiant instability phenomenon
which characterizes the composed Kerr-massive-scalar-field system,
we have explored here the possible existence of an analogous
superradiant instability for highly-charged Reissner-Nordstr\"om
black holes [in the regime $8/9<(Q/M)^2<1$] coupled to charged
massive scalar fields. It was shown that the two physical mechanisms
which are required to operate simultaneously in order to trigger the
superradiant instability phenomenon in the black-hole spacetime
[namely: (1) the superradiant amplification of the charged scalar
fields by the charged black hole, and (2) the existence of a binding
potential well in the black-hole exterior region which prevents the
extracted energy and electric charge from escaping to infinity] {\it
cannot} operate simultaneously. In particular, we have proved that
RN black holes in the charge interval $8/9<(Q/M)^2<1$ cannot support
bound-state resonance (with $\omega^2<\mu^2$) of charged massive
scalar fields in the superradiant regime $\omega<qQ/r_+$. This fact
suggests that the dynamics of charged massive scalar fields in these
highly-charged RN black-hole spacetimes is expected to be stable.

The stability results presented in this paper for the highly-charged
RN black holes in the regime $8/9<(Q/M)^2<1$, combined with former
analytical studies which explored the stability of these charged
black holes in the complementary regime ${(Q/M)}^2\leq 8/9$
\cite{Hodnbt,Notenbt2}, establish the fact that charged
Reissner-Nordstr\"om black holes are immune against the superradiant
instability phenomenon
in the entire parameter space.

Finally, we would like to stress that the present analysis excludes
the existence of exponentially growing instabilities (bound-state
resonances) of the charged scalar fields in the charged
Reissner-Nordstr\"om black-hole spacetime. However, it is worth
noting that milder forms of instabilities could appear, in principle
\cite{Noteref}.

\bigskip
\noindent
{\bf ACKNOWLEDGMENTS}
\bigskip

This research is supported by the Carmel Science Foundation. I thank
Yael Oren, Arbel M. Ongo, Ayelet B. Lata, and Alona B. Tea for
helpful discussions.



\begin{thebibliography}{99}

\bibitem{Zel} Ya. B. Zel'dovich, Pis`ma Zh. Eksp. Teor. Fiz. {\bf
14}, 270 (1971) [JETP Lett. {\bf 14}, 180 (1971)]; Zh. Eksp. Teor.
Fiz. {\bf 62}, 2076 (1972) [Sov. Phys. JETP {\bf 35}, 1085 (1972)];
A. V. Vilenkin, Phys. Lett. B {\bf 78}, 301 (1978).

\bibitem{PressTeu1} W. H. Press and S. A. Teukolsky, Nature {\bf
238}, 211 (1972); W. H. Press and S. A. Teukolsky, Astrophys. J.
{\bf 185}, 649 (1973).

\bibitem{Ins2} T. Damour, N. Deruelle and R. Ruffini, Lett. Nuovo Cimento {\bf
15}, 257 (1976); T. M. Zouros and D. M. Eardley, Annals of physics
{\bf 118}, 139 (1979); S. Detweiler, Phys. Rev. D {\bf 22}, 2323
(1980); H. Furuhashi and Y. Nambu, Prog. Theor. Phys. {\bf 112}, 983
(2004); V. Cardoso and J. P. S. Lemos, Phys. Lett. B {\bf 621}, 219
(2005); V. Cardoso and S. Yoshida, JHEP 0507:009 (2005); S. R.
Dolan, Phys. Rev. D {\bf 76}, 084001 (2007); S. Hod and O. Hod,
Phys. Rev. D {\bf 81}, Rapid communication 061502 (2010)
[arXiv:0910.0734]; H. R. Beyer, J. Math. Phys. {\bf 52}, 102502
(2011); Y. S. Myung, Phys. Rev. D {\bf 84}, 024048 (2011); S. Hod,
Phys. Lett. B {\bf 713}, 505 (2012); J. P. Lee, JHEP {\bf 1201}, 091
(2012); S. Hod, Phys. Lett. B {\bf 718}, 1489 (2013)
[arXiv:1304.6474]; R. Brito, V. Cardoso, and P. Pani, Phys. Rev. D
{\bf 88}, 023514 (2013); S. R. Dolan, Phys. Rev. D {\bf 87}, 124026
(2013); H. Witek, V. Cardoso, A. Ishibashi, and U. Sperhake, Phys.
Rev. D {\bf 87}, 043513 (2013); V. Cardoso, Gen. Relativ. and
Gravit. {\bf 45}, 2079 (2013); J. C. Degollado and C. A. R.
Herdeiro, Gen. Rel. Grav. {\bf 45}, 2483 (2013); R. Li, The Euro.
Phys. Journal C {\bf 73}, 2274 (2013); S. J. Zhang, B. Wang, E.
Abdalla, arXiv:1306.0932; H. Witek, arXiv:1307.1145; Y. S. Myung,
Phys. Rev. D {\bf 88}, 104017 (2013); R. Li, Phys. Rev. D {\bf 88},
127901 (2013); H. Okawa, H. Witek, and V. Cardoso, Phys. Rev. D {\bf
89}, 104032 (2014); B. Arderucio, arXiv:1404.3421; M. O. P. Sampaio,
C. Herdeiro, M. Wang, Phys. Rev. D {\bf 90}, 064004 (2014); Y.
Brihaye, C. Herdeiro, and E. Radu, Phys. Lett. B {\bf 739}, 1
(2014); C. L. Benone, L. C. B. Crispino, C. Herdeiro, and E. Radu,
arXiv:1409.1593; C. Herdeiro, E. Radu, and H. Runarsson,
arXiv:1409.2877.

\bibitem{Hodstat} S. Hod, Phys. Rev. D {\bf 86}, 104026 (2012)
[arXiv:1211.3202]; S. Hod, The Euro. Phys. Journal C {\bf 73}, 2378
(2013) [arXiv:1311.5298]; S. Hod, Phys. Rev. D {\bf 90}, 024051
(2014) [arXiv:1406.1179]; S. Hod, Phys. Lett. B {\bf 739}, 196
(2014) [arXiv:1411.2609]; C. A. R. Herdeiro and E. Radu, Phys. Rev.
Lett. {\bf 112}, 221101 (2014); C. A. R. Herdeiro and E. Radu, Phys.
Rev. D {\bf 89}, 124018 (2014); C. A. R. Herdeiro and E. Radu,
arXiv:1405.3696.

\bibitem{Hodbou} S. Hod, Phys. Lett. B {\bf 708}, 320 (2012)
[arXiv:1205.1872].

\bibitem{Hodcq} S. Hod, Submitted for publication in Class. Quant.
Grav. (2014).

\bibitem{Noteunt} We use natural units in which $G=c=\hbar=1$.

\bibitem{Notesat} It has recently been shown \cite{Hodcq} that the upper bound
(\ref{Eq1}) on the instability regime (that is, the upper bound on
the scalar field mass which can trigger the superradiant
instability) can be approached arbitrarily close in the eikonal
(large-$m$) limit \cite{Hodcq}.

\bibitem{Ins1} V. Cardoso, O. J. C. Dias, J. P. S. Lemos and S.
Yoshida, Phys. Rev. D {\bf 70}, 044039 (2004) [Erratum-ibid. D {\bf
70}, 049903 (2004)]; J. C. Degollado, C. A. R. Herdeiro, and H. F.
R\'unarsson, Phys. Rev. D {\bf 88}, 063003 (2013); J. C. Degollado
and C. A. R. Herdeiro, Phys. Rev. D {\bf 89}, 063005 (2014); S. Hod,
Phys. Rev. D {\bf 88}, 064055 (2013) [arXiv:1310.6101]; R. Li,
arXiv:1404.6309; S. Hod, Phys. Rev. D {\bf 88}, 124007 (2013)
[arXiv:1405.1045]; S. Hod, Phys. Lett. B {\bf 736}, 398 (2014); R.
Li and J. Zhao, The Euro. Phys. Journal C {\bf 74} 3051 (2014); S.
Hod, Phys. Lett. B {\bf 736}, 398 (2014) [arXiv:1412.6108]; C. L.
Benone, L. C. B. Crispino, C. Herdeiro, and E. Radu,
arXiv:1412.7278.

\bibitem{Bekch} J. D. Bekenstein, Phys. Rev. D {\bf 7}, 949 (1973).

\bibitem{NoteQ} We shall assume, without loss of generality, that
$Q\geq 0$.

\bibitem{Noteme} A superradiant instability which is expected
to be analogous to the superradiant instability of the Kerr black
hole described above.

\bibitem{Hodnbt} S. Hod, Phys. Lett. B {\bf 718}, 1489 (2013) [arXiv:1304.6474].

\bibitem{Notenbt2} The stability of {\it extremal} ($|Q|=M$) charged RN
black holes to charged scalar perturbations was proved in S. Hod,
Phys. Lett. B {\bf 713}, 505 (2012).

\bibitem{Chan} S. Chandrasekhar, {\it The Mathematical Theory of Black
Holes}, (Oxford University Press, New York, 1983).

\bibitem{HodPirpam} S. Hod and T. Piran, Phys. Rev. D {\bf 58},
024017 (1998) [arXiv:gr-qc/9712041]; S. Hod and T. Piran, Phys. Rev.
D {\bf 58}, 024018 (1998) [arXiv:gr-qc/9801001]; S. Hod and T.
Piran, Phys. Rev. D {\bf 58}, 024019 (1998) [arXiv:gr-qc/9801060].

\bibitem{Stro} T. Hartman, W. Song, and A. Strominger, JHEP 1003:118 (2010).

\bibitem{HodCQG2} S. Hod, Class. Quant. Grav. {\bf 23}, L23 (2006) [arXiv:gr-qc/0511047].

\bibitem{Hodpla} S. Hod, Phys. Lett. A {\bf 374}, 2901 (2010)
[arXiv:1006.4439].

\bibitem{Notemq} Note that $\mu$ and $q$ stand for $\mu/\hbar$ and $q/\hbar$, respectively.
Hence, these field parameters have the dimensions of
$($length$)^{-1}$.

\bibitem{Noteom} We shall henceforth omit the harmonic indexes $l$ and $m$ for brevity.

\bibitem{Hartle} J. B. Hartle and D. C. Wilkins, Commun. Math. Phys. {\bf 38}, 47 (1974).

\bibitem{Heun} A. Ronveaux, {\it Heun's differential equations}.
(Oxford University Press, Oxford, UK, 1995).

\bibitem{Flam} C. Flammer, {\it Spheroidal Wave Functions} (Stanford
University Press, Stanford, 1957).

\bibitem{Abram} M. Abramowitz and I. A. Stegun, {\it Handbook of
Mathematical Functions} (Dover Publications, New York, 1970).

\bibitem{Notehnb} The explicit form of the expansion coefficient $d$
is given in \cite{Hodnbt}. However, here we shall not need the
explicit form of this coefficient.

\bibitem{Noteaa} The argument for the positivity of the coefficient $a$ in the
frequency interval (\ref{Eq22}) goes as follows \cite{Hodnbt}: from
(\ref{Eq18}) one learns that the function $a(\omega)$ is in the form
of a convex parabola. Hence, $a(\omega)$ is minimized at the
boundaries of the frequency interval (\ref{Eq22}). Substituting
$\omega=0$ into (\ref{Eq18}), one finds $a(\omega=0)=M\mu^2>0$.
Substituting into (\ref{Eq18}) $\omega\to qQ/r_+$ for the case
$qQ/r_+\leq\mu$, one finds $a(\omega\to
qQ/r_+)=M\mu^2+{{Q^2q^2}\over{r_+}}(1-{{2M}\over{r_+}})>0$.
Substituting into (\ref{Eq18}) $\omega\to\mu$ for the case
$\mu<qQ/r_+$, one finds
$a(\omega\to\mu)=\mu(Qq-M\mu)>\mu^2(r_+-M)\geq 0$. One therefore
finds that, in the frequency interval (\ref{Eq22}), the coefficient
$a$ is positive definite in the entire range $0\leq (Q/M)^2\leq 1$.

\bibitem{Noteqa} It was proved in \cite{Hodnbt} that $c$ is
negative definite for charged RN black holes in the regime
$(Q/M)^2\leq 8/9$.

\bibitem{Notemuqq} Note that the two inequalities in (\ref{Eq29}) can
only be satisfied in the regime $\mu>qQ/2r_-$.

\bibitem{Notecon} This inequality follows from the fact that $I(\omega;Q/M)$ is in
the form of a convex parabola. Thus,
$I(x_2;Q/M)>\text{min}\{I(x_1;Q/M),I(x_3;Q/M)\}>0$ for
$x_1<x_2<x_3$.

\bibitem{Noteref} I would like to thank the anonymous referee for
this valuable comment.

\end{thebibliography}
\end{document}